# Scanning acoustic microscopy for mapping the microstructure of soft materials

N. G. Parker and M. J. W. Povey
*School of Food Science and Nutrition*
*University of Leeds, Leeds, LS2 9JT, United Kingdom*
n.g.parker@leeds.ac.uk

**Abstract**

Acoustics provides a powerful modality with which to 'see' the mechanical properties of a wide range of elastic materials. It is particularly adept at probing soft materials where excellent contrast and propagation distance can be achieved. We have constructed a scanning acoustic microscope capable of mapping the physical microstructure of such materials. We review the general principles of scanning acoustic microscopy and present new examples of its application in imaging biological matter, industrial materials and particulate systems.

## 1. Introduction

The propensity of acoustics to propagate, and thereby probe, within many media, along with its non-invasive nature, has led to major applications, e.g, in non-destructive testing of materials (Krautkrämer, 1990; Kundu, 2003), imaging within human tissue (Duck, 1998) and monitoring of industrial processes (Povey, 1997). It provides information on the elastic properties of the media, and thus provides distinct information to other imaging modalities such as light. Furthermore, ultrasound can probe within optically opaque materials. The scanning acoustic microscope employs ultrasound to 'see' the mechanical properties of materials at high resolution and was first demonstrated by Lemons and Quate (1974). An ultrasound beam insonifies a point in a sample and the received signal indicates the geometry and composition of that region. By scanning the ultrasound beam around in space, a three-dimensional acoustic image of the material is obtained. From this important mechanical properties such as elastic moduli, density and compressibility can typically be inferred. Resolution lengthscales of as low as 15nm have been reported in cryogenic environments (Moulthrop, 1992). For the more typical case where water is used as the coupling environment, the resolution lies within the range 400nm-200µm (Briggs, 1992). At its lower end this surpasses optical resolution, and the images themselves often carry more information than other imaging modalities. For these reasons, scanning acoustic microscopy is an attractive and powerful tool, and has already had success in mapping the micro- and nano-structure of construction materials, semiconductor devices and biological specimens (Kundu, 2003).

Ultrasound is particularly well suited to imaging soft materials, such as colloids, biological specimens, polymers and liquids. These materials typically offer both good acoustic contrast and penetration depth, which enables detailed sub-surface imaging. We have constructed a scanning acoustic microscope for the imaging and characterization of such soft materials down to the microscale. We will briefly review the basic principles of ultrasound propagation and imaging, before discuss the scanning acoustic microscope that we have developed. We will then present results from a range of systems and analyse them qualitatively. These will serve to demonstrate the wide possibilities offered by the scanning acoustic microscope.

## 2. Basic principles of acoustic propagation

Ultrasound waves are sound waves with a frequency above the limit of human hearing (approximately 20 kHz). Ultrasound can propagate deep within many materials and is capable of resolving fine features. In general sound can propagate as both longitudinal and shear forms. However, in fluids, where shear effects are limited to boundary layers, longitudinal waves dominant the bulk material. This consists of periodic compressions and rarefactions in space which occur in the direction of propagation, i.e., a pressure wave.

The ultrasound beam is sensitive to several factors within the medium. The propensity of a medium to support sound waves is parameterized by its characteristic impedance $Z$, which is given by $Z=\rho c$ where $\rho$ is the density of the medium and $c$ is the speed of sound. When a sound wave encounters a discontinuity in $Z$, it is partially transmitted and partially reflected. For example, for plane waves at normal incidence to an interface between two media (denoted 1 and 2), the reflection coefficient (ratio of the reflected amplitude to the incident amplitude) is,

$$R=|Z_1-Z_2|/(Z_1+Z_2). \qquad (1)$$

For waves at oblique incidence the situation becomes complicated by a strong angular dependence, generation of surface waves and mode conversion between shear and longitudinal waves (Krautkrämer, 1990). However, these effects are well understood and can be taken into account. The reflected signals thus indicate the boundaries in the medium, i.e. the surface topography and sub-surface tomography. Importantly, key mechanical parameters can also be determined from the returning signal providing a suitable model can be applied. For example, in liquids the acoustic impedance is given by,

$$Z=(\rho K)^{1/2}. \qquad (2)$$

Providing the density can be measured then the bulk modulus $K$ can be derived. Similarly, for solids, the acoustic impedance is given by,

$$Z=[\rho(K+4/3\ G)]^{1/2}, \qquad (3)$$

where the shear modulus $G$ can additionally be inferred.

As the beam propagates through the sample it will also be scattered by small inhomogeneities and lose energy by the transfer of sound energy into heat. This causes the pressure amplitude of the wave to attenuate with position according to $p(x)=p_0\exp(-\alpha x)$, where $p_0$ is the unattenuated amplitude and $\alpha$ is the attenuation coefficient. Attenuation measurements can therefore be used to infer the absorption and scattering properties of the medium. For example, this has been used to great effect in fluids to determine the particle size distribution and viscoelastic quantities (Povey, 1997).

## 3. Scanning acoustic microscope

Figure 1(a) presents a schematic of the scanning acoustic microscope. The central component is a high frequency piezoelectric transducer unit (Panametrics V3534). The transducer is immersed within a coupling fluid, for which we employ distilled water. (Note that, where fluids are to be imaged, the coupling fluid may be the sample itself.) Variations in temperature can modify the acoustic properties of the coupling system, e.g., the speed of sound, and significantly distort the image. Consequently an external temperature bath (not shown) is essential for high resolution imaging. A broadband radio frequency pulse (UTEX 320 pulse generation unit) excites the piezoelectric crystal into generating a burst of ultrasonic waves with dominant frequency $f$ = 63 MHz. A damping layer ensures that effectively only one pulse is produced which propagates along a quartz delay rod towards the sample. A spherical lens ground into the base of the delay rod refracts the acoustic beam towards a focal point in the coupling medium at a distance of $F$ = 6 mm. The transducer is operated in reflection mode, where it acts as both the emitter and receiver of the ultrasound. The ultrasound which is reflected and/or scattered back into the transducer excites the piezoelectric crystal, generating a voltage which is detected by an oscilloscope. Typically some averaging is performed to improve the signal-to-noise ratio. Note that the dynamic range of our current system is 70dB. The signal is subsequently exported to a computer where it is processed and visualized.

A significant feature of the acoustic microscope is that spherical and chromatic aberrations are typically negligible such that the lateral resolution corresponds to the diffraction limit (Briggs, 1992). According to the Rayleigh criterion this resolution is given by,

$$R_L = 0.6Fc/fD, \qquad (4)$$

where $D$ is the lens diameter and $c$ is the speed of sound. For our microscope ($D$ = 6 mm and $c$ = 1540 ms$^{-1}$) this gives $R_L \approx 15\mu m$. We have confirmed this experimentally by imaging a sharp edge and measuring the spread in the returning image. Furthermore, the pulse length in water is approximately $50\mu m$ (full-width-half-maximum of the pulse). The basic imaging resolution of our system is therefore $15 \times 15 \times 50 \ \mu m$. Improved resolution is possible with higher frequency transducers, as illustrated in Figure 1(b). However, this comes at the expense of the working distance since the attenuation of water (and indeed most fluids) over the frequencies of interest increases with the square of the frequency. For example, a 2 GHz transducer has a resolution well below 1 micron but a restrictive propagation depth of about 60$\mu$m. For our current microscope the working distance is over 1cm, providing good all-round capabilities. Note that the working distance can be extended by working at raised temperature since the attenuation in water decreases with temperature ($\alpha$ decreases by a factor of 2 from 20 $^0$C to 50 $^0$C) while resolution can also be improved at a factor of $\sqrt{2}$ by operating in the nonlinear acoustic regime (Rugar, 1984).

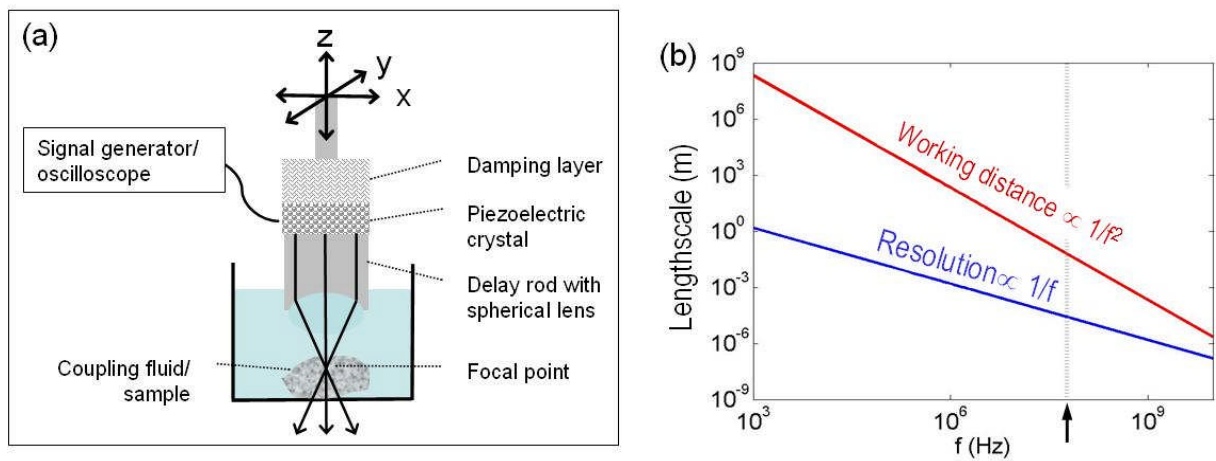

**Figure 1: (a) The scanning acoustic microscope. A piezoelectric crystal generates a sound pulse which propagates along the delay rod (back-propagating waves are heavily damped by an adjacent damping layer). A spherical lens focuses the beam within the sample. The reflected signal follows the return journey, inducing a voltage which is detected by an oscilloscope. The transducer unit is scanned in three-dimensions as required. (b) Imaging resolution and propagation depth in water at 20 $^0$C as a function of frequency. We define the propagation depth as the lengthscale over which the pressure amplitude decreases to 1% of its original value (a loss of 40dB). The arrow indicates the parameters of the current system.**

The transducer unit is mounted on a three-dimensional positioning system accurate to a few microns. This error arises primarily from mechanical vibrations and inaccuracies in the stepper motors that drive the system. Various scanning modes are possible. A 2D scan in the *x-y* plane is sufficient to establish a 3D image since axial information is provided by the time-of-flight of the acoustic beam. Although the beam diameter, and thus the resolution, decreases with distance from the focal point, providing the region of interest is no deeper than 200 microns a resolution of 50 microns is possible. We will employ this approach in this paper. In particular we will present surface topography plots (where the position of the first reflection is mapped) and C-scans (slices through the sample at fixed time-of-flight/depth). Note that for thicker samples and where maximum resolution is required throughout, the focal point itself should be moved in space for true 3D mapping.

A final important consideration is the angular reflection from the sample. For a surface which is normal to the incident beam the waves are retro-reflected back to the transducer. Tilting the surface results in angular reflection which reduces the proportion of the reflected energy that falls on the transducer. In this manner the surface topography modulates the returning acoustic amplitude. From simple geometrical arguments we expect that the received signal becomes negligible when the surface is tilted away from the horizontal by around $20^0$ and this is consistent with what we observe experimentally.

## 4. Case studies

Having described the principles of the scanning acoustic microscope we will now present some results. We will consider five distinct systems that demonstrate the wide capabilities of acoustic microscopy. From the ensuing acoustic images it is typically possible to map quantitatively the mechanical parameters of the system, such as the elastic moduli and density, as indicated in Section 2. Quantitative models of scanning acoustic microscopy, which relate the signal to these mechanical parameters, can be found elsewhere for a range of systems (Briggs, 1992; Kundu, 2003). This, however, is beyond the scope of this current work and so our analysis is limited to a qualitative discussion of the images.

### *4.1 Biological cells*

One of the most exciting applications of the acoustic microscope is in imaging biological cells (Kundu, 2003). Here its non-destructive nature is of paramount importance. Although this was recognized in the pioneering work on the acoustic microscope (Lemons, 1974; Hildebrand, 1981) it has been, somewhat surprisingly, underused as a means to probe cells. To date, acoustic microscopy has focused on cells derived from humans and animals. These cells are sufficiently small that the returning signal features strong interference between the reflections from the front cell wall, the back cell wall, and the substrate upon which the cell is fixed. Furthermore, the reflection from the substrate (typically a glass slide) tends to dominate the reflection. Although this complicates the analysis theoretical techniques have now been developed to overcome this, e.g., by treating the cellular reflection as a perturbation to the reflection from the substrate (Kundu, 1991; Kundu, 2000).

We will here consider plant cells. Plant cells can be sufficiently large (typically an order of magnitude larger than human/animal cells) that the acoustic signals from the front and back of the cell are resolvable in time, thus negating undesirable interference effects. We will specifically consider onion skin. This single layer membrane has the added advantage that is can be stretched over a frame, thus removing the presence of a strongly reflecting substrate.

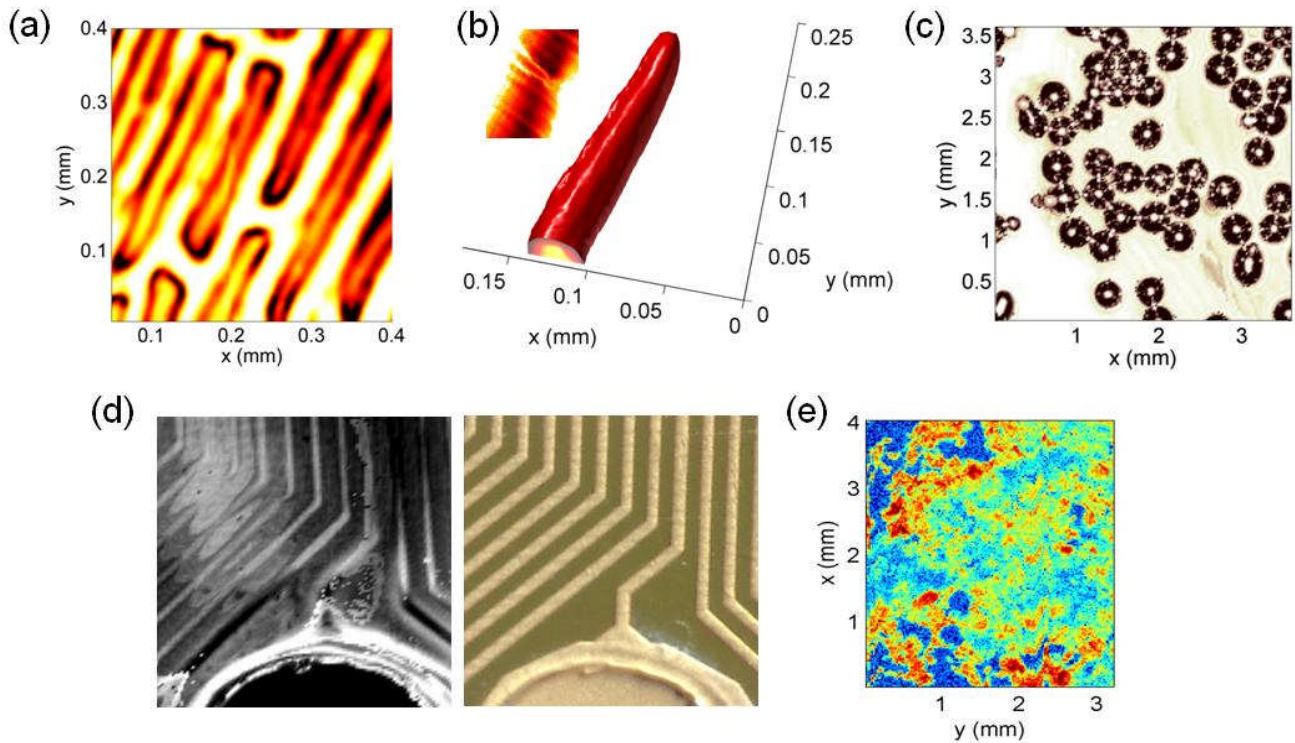

**Figure 2: Images from the scanning acoustic microscope. (a) A membrane of onion cells. (b) A human hair shaft and a fractured hair (inset). (c) Glass microspheres at a liquid-solid interface. (d) Acoustic (left) and optical (right) images of electronic circuitry. (e) Polymer foam scaffold for use in tissue engineering. In (a), (c) and (d) light/dark areas represent high/low signal, in (b, inset) light/dark areas correspond to low/high signal, and in (e) red/blue correspond to high/low signal.**

Figure 2 (a) presents a C-scan acoustic image of the onion cells (image of a plane at fixed depth in the sample). The images show remarkable clarity. The membrane exhibits strong acoustic contrast while the $20\mu m$ resolution imaging limit is clearly sufficient to map the macroscopic cellular structure. With greater resolution it is possible to reveal sub-cellular features such as organelles (Kundu, 2003). The cell topography modulates the image with retro-reflecting surfaces, i.e., the cell centre and walls, providing the brightest signal. Notably, the cell wall is brighter than the cell centre, suggesting a larger impedance than the rest of the cell.

For cells it is relatively straightforward to convert the reflected signal into mechanical parameters since the cell can be adequately modeled as a layer of elastic fluid, for which the acoustic properties are well known. Detailed discussion of this model can be found in the works of Briggs (1992) and Kundu (2003).

*4.2    Hairs/fibres*

Hair is an interesting physical system due to its complex fibrous structure and remarkable strength. It also plays an important acoustic function in nature. In the inner ear tiny hairs are responsible for detecting sound waves while head hair is thought to play a role in human auditory localization (Treeby, 2007). Despite this the acoustic properties of hair are not well documented. Furthermore, the acoustic properties of polymer fibers as a whole are important for sound absorption applications. Figure 2(b) presents an acoustic microscope image of the surface of a human hair. A strong and clear signal is returned. Narrow fractures are also revealed (inset). The scanning acoustic microscope has the potential to accurately determine the acoustic properties of single hair/polymer fibres. In contrast, the acoustic properties of hair have been determined to date from collective samples (Treeby, 2007).

*4.3    Particulate systems*

Particulate systems occur ubiquitously in industry. When the particle size exceeds the resolution limit acoustic microscopy can be employed to spatially image the structure and properties of the particles, as well as their spatial distribution. Figure 2(e) presents an image of glass microspheres at a water-glass interface. The spheres are approximately 400 microns in diameter and are mostly regular, although ellipsoidal particles are also evident. The acoustic signal from each sphere consists of a central bright spot and a dark periphery. Assuming the particles to be spherical the surface angle at the bright-dark transition point is given by $\theta = \sin^{-1}(a/r)$, where $r$ is the particle radius and $a$ is the radius of the bright spot. This typically corresponds to around $16^0$ for the spheres presented, which agrees well with the critical reflection angle discussed in Section 3. Also evident from the image is the striking series of bright lines that exist between neighbouring particles. This is thought to be due to constructive interference of the angularly reflected waves from adjacent spheres or due to modes of the spheres themselves. This will be analysed in a future work.

When the imaging resolution exceeds the particle size scattering dominates the acoustic propagation. The scattering is highly sensitive to the average size and size distribution of the particles. Spectral attenuation and velocity measurements are commonly employed to infer these statistical parameters (Povey, 1997). Using the acoustic microscope it is feasible to map out the mean particle size and particle size distribution throughout a sample. This could, for example, be used to study creaming, sedimentation and aggregration in small systems.

*4.4    Foams*

Acoustic microscopy is also suited to the analysis of foams where a detailed mapping/understanding of the structure is required. One example are the biodegradable polymer foams used as scaffolds for tissue engineering (Langer, 2003). A controlled pore structure is essential to mimic the extracellular structure of the native tissue and provide pathways for the transport of nutrients and waste. Current scaffold imaging relies on x-ray computed tomography, scanning electron microscopy and magnetic resonance imaging (Mather, 2008). Acoustic microscopy provides a new modality for imaging tissue scaffolds with a key benefit that it can be employed during tissue growth. Figure 2(c) presents a C-scan through a polymer foam scaffold. The irregular image arises from the irregularity and roughness of the scaffold structure. The size and distribution of pores can be readily determined from such images. Similar to the particulate systems, when the features become smaller than the resolution limit spectral analysis of velocity and attenuation measurements may provide an approach to statistically map out the pore structure of the foam (Roth, 1995).

*4.5    Electronic/integrated circuits*

The potential of the acoustic microscope to map the subsurface mechanical properties of a material has found strong applications in non-destructive testing of materials (Kräutkramer, 1990; Kundu, 2003). One subset of this is in electronic/integrated circuits (Lemons, 1974).

Figure 2(d) presents a comparison between the acoustic and optical image of an electronic circuit. It is important to note that a plastic coating exists over the circuitry. If this was an optically opaque layer instead then only the acoustic image would reveal the underlying structure. It is clear that the acoustic image contains greater details than the optical image, as noted previously (Lemons, 1974). This is likely to arise from either acoustic-sensitive aberrations in the coating or subsurface details.

## 5. Conclusions

The scanning acoustic microscope provides a powerful method to image the microstructure and even nanostructure of materials. Importantly, acoustics probes the mechanical/elastic properties of the system, and is thus distinct from the many other imaging techniques available. Using our current microscope we have demonstrated high contrast and high resolution imaging of biological cells, hairs/fibres, materials, foams and particulate systems. In the future we will extend our analysis from a qualitative level to provide detailed quantitative information of the mechanical properties of these systems.